\documentclass[journal]{IEEEtran}
\usepackage{tikz}
\usepackage{pgfplots}
\usepackage{pgfplotstable}
\usepgfplotslibrary{fillbetween, patchplots, polar}
\tikzset{>=latex}
\usetikzlibrary{
  3d,
  arrows,
  arrows.meta,
  calc,
  matrix,
  plotmarks,
  positioning,
  shadings,
  shapes.geometric,
  decorations.pathmorphing
}
\tikzset{
  currentarrow/.style={
    -{Stealth[length=2.2mm,width=0.8mm]},
    line width=0.6pt,
    line cap=round
  }
}

\pgfplotsset{compat=newest}
\usepackage{standalone}    % 
\newlength\figurewidth
\newlength\figureheight
\usepackage{amsmath,amssymb,amsfonts}
\usepackage{graphicx} 
\usepackage{cite}
\usepackage{xcolor}
\newcommand{\tM}[1]{\widetilde{\mathbf{#1}}}

\providecommand*{\M}[1]
{\mathbf#1}        % matrices
\providecommand*{\V}[1]{\boldsymbol#1} % (physical) vectors
\providecommand*{\UV}[1]{\hat{\boldsymbol#1}}  % unit vectors
\providecommand*{\tV}[1]{\tilde{\boldsymbol#1}}
\providecommand*{\T}[1]{\mathrm{#1}}  % text 
\providecommand*{\ju}{\ensuremath{\T{j}}} % j
\providecommand*{\diff}{\operatorname{d}\!}  % d in integrals
\newcommand{\R}{\mathbb{R}{}}

\newcommand{\regT}{\varOmega_\T{T}}
\newcommand{\regR}{\varOmega_\T{R}}

\begin{document}

\newpage
\pagestyle{headings}
%\twocolumn

\title{Near-Field Sampling for Line Sources}

\author{Jiawang Li, Mats Gustafsson

\thanks{Manuscript received \today. This work was supported in part by a Project Grant in ELLIIT 
Call  D, in  part  by  NextG2Com  (grant  no.  2023-00541) 
funded by the VINNOVA program for Advanced Digitalisation, and in part by Swedish Research Council SEE-6GIA 2024-06482.  (Corresponding author: \textit{Jiawang Li}).}%
\thanks{Jiawang Li and Mats Gustafsson are with the Department of Electrical and Information Technology, Lund University, 22100 Lund, Sweden (e-mail: {\{jiawang.li, mats.gustafsson\}}@eit.lth.se).
}%
}

\maketitle

\begin{abstract}
Near-field sampling seeks to represent electromagnetic fields between transmitting and receiving regions using a minimal number of measurement points while preserving the dominant spatial modes. This paper develops a geometry-aware sampling framework based on spatial degrees of freedom (DoF). A view-length formulation is used to derive closed-form expressions for the propagating-mode DoF density for simple line-source geometries, providing both the total DoF and its local distribution. One-DoF sampling points are obtained from equal increments of the cumulative DoF density, yielding an adaptive nonuniform sampling strategy up to the knee of the singular-value spectrum. To improve the representation of the remaining modes beyond the knee, a reactive-mode density is introduced to guide the placement of additional edge samples. An operator-based sampling error functional is formulated and shown to be lower-bounded by the neglected singular values of the continuous channel operator. Numerical results demonstrate that the proposed sampling strategy closely approaches the optimal performance obtained from singular-value decomposition and significantly outperforms sampling based solely on the propagating-mode DoF density. 
\end{abstract}


\section{Introduction}
\IEEEPARstart{S}{patial} sampling of electromagnetic fields is fundamental to antenna characterization~\cite{qureshi2012efficient, mezieres2021antenna,fuchs2017fast}, imaging problems ~\cite{capozzoli2016singular,randazzo2021two}, and inverse source problems~\cite{maisto2021near,marengo2000inverse,han2026reactive,solimene2018inverse,bucci1997electromagnetic}. For conventional planar field measurements, uniformly distributed sampling points are typically arranged with a spatial spacing of approximately half a wavelength according to the classical sampling criterion~\cite{yaghjian1986overview,marks1993advanced}. Although such methods are simple and broadly applicable, they do not fully account for the geometries of the source and observation regions and may therefore require more samples than the number of independent spatial modes supported by the propagation channel~\cite{miller2000communicating,gabor1961light,bucci1989degrees}. This motivates sampling strategies that adapt the sampling distribution to the spatial information carried by the electromagnetic field.

Electromagnetic degrees of freedom (DoF) characterize the number of independent spatial field modes supported between source and observation regions in free space~\cite{franceschetti2017wave,
Bucci2025,miller2000communicating,poon2005degrees,
dardari2020communicating,jensen2008capacity,
puggelli2025maximizing,yuan2024breaking}. From a sampling perspective, the DoF represent the effective dimension of the spatial field space and provide a reference for the number of samples required for accurate field representation. Existing DoF studies include Weyl's law~\cite{weyl1911asymptotische}, nonredundant field representations~\cite{bucci1989degrees,bucci1998representation}, singular-value analysis of propagation operators~\cite{miller2000communicating,poon2005degrees}, and paraxial approximation~\cite{miller2019waves,piestun2000electromagnetic}. More recently, the mutual-shadow approach has enabled the DoF and the spectral transition of electrically large propagation systems to be estimated directly from geometric quantities~\cite{gustafsson2025shadow,gustafsson2025degrees}. However, existing results mainly characterize the total DoF, indicating how many samples may be required but not where they should be placed. The local distribution of DoF and its use for geometry-aware sampling remain less explored.

Beyond uniform half-wavelength sampling, nonredundant methods reduce field measurements by exploiting the geometry-dependent spatial bandwidth~\cite{bucci1998representation}. In these methods, the field is represented in a geometry-adapted coordinate system, allowing the sampling rate to follow the local field variation rather than a fixed physical spacing. Later works used geometry-dependent parameterizations and warping transformations to equalize the spatial bandwidth and construct compact sampling grids for planar, cylindrical, and more general observation surfaces~\cite{maisto2021efficient,maisto2021near}. Other approaches determine the effective dimension of the radiation operator from its asymptotic behavior or singular-value structure and construct discretizations that preserve the dominant modal content~\cite{pierri2020asymptotic,pierri2021ndf,leone2022dimension,solimene2019sampling,migliore2025intuitive}. Compressed sensing method can further reduce the number of measurements when the field or source distribution is sparse or compressible in a prescribed representation domain~\cite{candes2008introduction,bangun2022optimizing}. Reducing redundancy requires determining not only the minimum number of samples needed to preserve the relevant field information, but also their appropriate spatial locations. Although these techniques can substantially reduce sampling redundancy, they are often tailored to particular geometries, require an explicit parameterization of the observation domain, or depend on prior knowledge of the spectral properties of the propagation operator.

In this paper, we develop a view-length-based framework for geometry-aware near-field sampling. Instead of characterizing only the total number of DoF (NDoF), the proposed formulation introduces a local DoF density that describes how the available propagating modes are distributed along the receiving line. This provides a direct geometric interpretation of spatial-mode accumulation and establishes an explicit link between the source--receiver geometry and the placement of sampling points. Closed-form expressions are derived for line-source and line-receiver geometries embedded in three-dimensional (3D) space, yielding both the total DoF and its spatial distribution.

The sampling strategy combines the propagating-mode DoF density with an additional reactive-mode density that accounts for the localized modal behavior beyond the spectral knee. In the propagating region, sampling points are obtained from equal increments of the cumulative DoF density, producing one-DoF sampling intervals and revealing a close connection to the reduced-field formulation of Bucci \textit{et~al.}~\cite{bucci1998representation}. While both approaches accurately describe the dominant propagating modes, they become increasingly suboptimal beyond the spectral knee, where reactive modes exhibit stronger edge localization~\cite{brick2026interpreting}. To capture these reactive modes, the proposed framework introduces additional geometry-dependent sampling points guided by the reactive-mode density. An operator-based sampling error functional is derived and shown to be lower-bounded by the first neglected singular value of the continuous channel operator. Numerical examples demonstrate that the proposed propagating--reactive sampling strategy closely reproduces the singular-value spectrum and significantly improves post-knee performance compared with sampling based solely on the propagating-mode density.

The rest of this paper is organized as follows. Section~II introduces the propagation model and formulates the sampling error. Section~III derives the view-length-based DoF density and develops DoF-edge sampling together with a reactive-mode correction. Section~IV validates the proposed sampling framework for representative non-coplanar and curved Tx--Rx configurations. Finally, Section~V concludes the paper.

\textit{Notation:} Throughout this paper, boldface letters indicate vectors and boldface uppercase letters designate matrices. Superscript $(\cdot)^{\T{H}}$ and $(\cdot)^{\dagger}$ stand for Hermitian transpose and pseudo-inverse~\cite{ben2003generalized}.

\section{Modelling and error formulation}
\begin{figure}[t]
  \centering
  \includegraphics[width=1\linewidth]{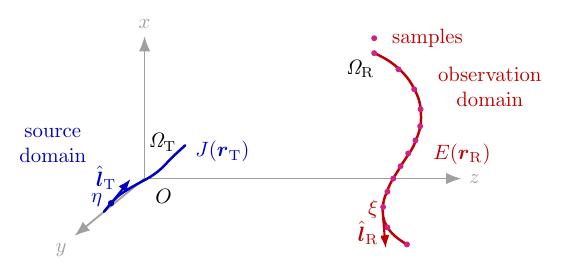}
 \caption{General geometry for the sampling problem. The transmitting region $\it\Omega_{\T T}$ is represented by a finite line source, while the receiving region $\it\Omega_{\T R}$ is an gently curved line embedded in 3D space, with $\hat{\boldsymbol l}_{\T T}$ and
$\hat{\boldsymbol l}_{\T R}$ denoting their respective local unit
tangent vectors.}  
\label{fig:research_problem}
\end{figure}

Consider a transmitting region $\regT$ and an observation region $\regR$, as illustrated in Fig.~\ref{fig:research_problem}.
The transmitting region is represented by a line source distribution
$J(\V r_\T T)$ supported on $\regT$, while the radiated
field is observed over a line $\regR$.

The geometry is intentionally kept general. As illustrated in Fig.~\ref{fig:research_problem}, the observation region may be either a straight line or a gently curved line in 3D space, provided that its local tangent direction varies only gradually. Therefore, both conventional coplanar 2D configurations and non-coplanar 3D geometries are included.

For simplicity, we adopt the commonly used scalar propagation model~\cite{miller2000communicating,maisto2021efficient,leone2022dimension,Bucci2025}, which enables the geometry-dependent spatial information of the propagation channel to be analyzed without the additional complexity introduced by polarization coupling and multiple field components. This formulation facilitates the derivation and interpretation of the proposed DoF-density-based sampling strategy and provides a reference for the vector electromagnetic extension.
Let $\V r_\T T$ denote a source point on $\regT$ and
$\V r_\T R$ an observation point on $\regR$.
The radiated field can be expressed through the linear propagation
operator
\begin{equation}
E(\V r_\T R)=\int_{\it\Omega_{\T T}}
G(\V r_\T R,\V r_\T T) J(\V r_\T T) \diff l_\T T,
\label{eq:field_operator_general}
\end{equation}
where $G(\V r_\T R,\V r_\T T)={\T{exp}({-\ju kR})}/
({4\pi R)}$ denotes the free-space Green's function with \( R=\lvert\V{r_\T R}- \V{r_\T T}\rvert\) for electric current in $\mathbb{R}^3$~\cite{harrington1993field}, \(k=2\pi/\lambda\) is the free-space wavenumber and the \(\lambda\) is the corresponding wavelength. \(J(\V r_\T T)\) is the corresponding electric current. It establishes a linear mapping from the source space defined on $\regT$ to the observable field space defined on $\regR$.

A central problem in spatial field sampling is to determine how many independent spatial modes need to be sampled over the receiving region and where the corresponding sampling points should be placed.
The classical Shannon--Nyquist sampling theorem is formulated for square-integrable bandlimited signals~\cite{franceschetti2017wave,marks1993advanced}, where bandlimitedness is expressed as a constraint on the support of the signal's Fourier transform. In the present work, an analogous viewpoint is adopted. Rather than constraining the support of the Fourier spectrum, the sources are assumed to be confined to a finite transmitting region $\regT$, and the radiation operator~\eqref{eq:field_operator_general} mapping sources in $\regT$ to fields in the observation region $\regR$ replaces the role of the Fourier transform.

This viewpoint naturally leads to a formulation of sampling as an inverse source problem. Given a finite set of field samples in $\regR$, the objective is to estimate the source distribution in $\regT$ and subsequently reconstruct the field throughout $\regR$. The quality of a sampling scheme is therefore determined by how accurately the selected measurements preserve the source subspace responsible for the radiated field.

For notational simplicity, fields and currents are expanded in basis functions~\cite{harrington1993field} chosen such that their least-squares norms are preserved. Let the densely sampled field in $\regR$ be represented as
\begin{equation}
\M{E}_\T{R}=\M{H}\M{I}_\T{T},
\label{eq:dense_field}
\end{equation}
where $\M{H}$ is the densely sampled radiation operator~\eqref{eq:field_operator_general} and $\M{I}_\T{T}$ denotes the source coefficient vector. Selecting $N$ measurement points in $\regR$ (see in Fig.~\ref{fig:research_problem}) corresponds to applying a sampling operator $\M{P}_N$, yielding the sampled operator $\tM{H}=\M{P}_N\M{H}$ and the sampled measurements
\begin{equation}
\tM{E}_\T{R}=\tM{H}\M{I}_\T{T}.
\end{equation}

Assuming that only the sampled measurements are available, the source coefficients are estimated via the pseudo-inverse~\cite{ben2003generalized}
\begin{equation}
\tM{I}_\T{T} = \tM{H}^{\dagger}\tM{E}_\T{R} = \tM{H}^{\dagger}\tM{H}\M{I}_\T{T}.
\label{eq:estimated_source}
\end{equation}
The operator $\tM{H}^{\dagger}\tM{H}$ is the orthogonal projector onto the source subspace observable through the selected measurements.
The densely sampled field is then reconstructed as
\begin{equation}
\widehat{\M{E}}_\T{R} = \M{H}\tM{I}_\T{T} = \M{H}\tM{H}^{\dagger}\tM{E}_\T{R},
\label{eq:reconstructed_field}
\end{equation}
where $\M{H}\tM{H}^{\dagger}$ acts as an interpolation operator from the sampled measurements to the dense field representation.

The reconstruction error is therefore
\begin{equation}
\widehat{\M{E}}_\T{R}-\M{E}_\T{R} =\M{H} \big(\tM{H}^{\dagger}\tM{H}-\M{1}\big)\M{I}_\T{T},
\end{equation}
and the worst-case error over all unit \(L^2\)-norm source distributions defines the sampling error
\begin{equation}
\mathcal{E}(\M{P}_N)=\max_{\lVert \M{I}_\T{T} \rVert=1}
\bigl\lVert \widehat{\M{E}}_\T{R}-\M{E}_\T{R} \bigr\rVert
=\bigl\lVert\M{H}\big(\tM{H}^{\dagger}\tM{H}-\M{1}\big)\bigr\rVert_2,
\label{eq:errornorm}
\end{equation}
which equals the largest singular value of the error operator.

To interpret the sampling error in~\eqref{eq:errornorm}, we consider the reduced SVDs of the dense and sampled channel matrices, \(\M{H}=\M{U}\V{\Sigma}\M{V}^{\T{H}}\) and \(\tM{H}=\tM{U}\tV{\Sigma}\tM{V}^{\T{H}}\). The corresponding nonzero singular values are denoted by \(\sigma_n\) and \(\tilde{\sigma}_n\), respectively, and are arranged in nonincreasing order. The sampling error then reduces to 
\begin{equation}
\mathcal{E}(\M{P}_N)
=\bigl\lVert\V{\Sigma}\big(\M{V}^{\T H}\tM{V}\tM{V}^{\T H}\M{V}-\M{1}\big)\bigr\rVert_2\geq \sigma_{N+1},
\label{eq:error_bound}
\end{equation}
which is bounded from below by the singular value $\sigma_{N+1}$ of the continuous channel according to the Eckart–Young theorem~\cite{eckart1936approximation}.
This expression shows that the reconstruction accuracy is governed by the extent to which the sampled operator preserves the dominant right-singular subspace of the full radiation operator. In particular, accurate reconstruction requires the dominant source modes represented by the columns of $\M{V}$ to remain observable through the sampled operator $\tM{H}$. The residual error is therefore determined by the source subspace that is not captured by the selected measurements and is closely related to the neglected singular values of the radiation operator.

The sampling problem can therefore be interpreted as selecting as few receiver locations as possible while keeping the sampling error
\(\mathcal{E}(\M{P}_N)\) below a prescribed tolerance \(\delta\), where \(\M{P}_N\) denotes a row-selection matrix that extracts \(N\) rows from the dense channel matrix corresponding to \(M\) candidate sampling locations. Each selected row corresponds to one receiver sample, and distinct rows correspond to distinct sampling locations. The quantity \(\mathcal{E}(\M{P}_N)\) characterizes the accuracy loss caused by using the selected sampling locations instead of the dense sampling grid.

This sampling requirement is directly connected to the mutual-shadow (view) DoF estimate~\cite{gustafsson2025shadow}. The mutual-shadow (view) length or area provides a geometry-based estimation of the number of significant channel modes, and therefore gives an estimate of
the minimum number of sampling locations required before the singular values become small. In other words, the mutual-shadow (view) DoF indicates the sampling level at which the error is expected to approach the neglected singular-value level \(\sigma_{N+1}\).

\section{Sampling for scalar line sources and fields}\label{sec:3D_space}

We consider transmitting and receiving line regions whose geometric coupling is quantified by the view length~\cite{gustafsson2025shadow}. Assuming full mutual visibility without occlusion, the view length is determined by the tangent directions of the two lines and their relative separation.

For 2D and in-plane line geometries, the mutual shadow (or view) length~\cite{gustafsson2025shadow} is known to be
\begin{equation}    L_\T{TR}=\int_{\regT}\int_{\regR}\frac{\big|\UV{n}^\prime\cdot\V{R}\big|\,\big|\UV{n}\cdot\V{R}\big|}{|\V{R}|^3}\diff l_\T{R}\diff l_\T{T},
    \label{eq:2D_mutual_shadow_length}
\end{equation}
where \(\UV{n}^\prime\) and \(\UV{n}\) are the unit normals of the transmitting and receiving lines in the plane, respectively, and
\(\V{R}=\V{r}_\T{T}-\V{r}_\T{R}\).
The corresponding asymptotic number of spatial DoF per polarization is
\begin{equation}
    \mathcal N_\T{DoF}=\frac{L_\T{TR}}{\lambda}.
    \label{eq:2D_NDoF}
\end{equation}

For gently curved line in \(\R^3\) as depicted in Fig.~\ref{fig:research_problem}, the in-plane expression in
\eqref{eq:2D_mutual_shadow_length} can be generalized by taking into account the local angular properties of both the transmitting and receiving curves. Specifically, the view length is written as
\begin{equation}    L_\T{TR}=\int_{\regT}\int_{\regR}\frac{\big|\UV{l}_\T{T}\times\V{R}
\cdot\UV{l}_\T{R}\times\V{R}\big|}{|\V{R}|^3}\diff l_\T{R}\diff l_\T{T},
\label{eq:3D_NDoF}
\end{equation}
where \(\UV{l}_\T{T}\) and \(\UV{l}_\T{R}\) are the unit tangent vectors of the transmitting and receiving lines, respectively (see Fig.~\ref{fig:research_problem}). The numerator in \eqref{eq:3D_NDoF} accounts for the relative angular projections of the two line elements with respect to their separation vector. It can also be rewritten using the vector identity
\begin{equation}    \UV{l}_\T{T}\times\V{R}\cdot\UV{l}_\T{R}\times\V{R}=\UV{l}_\T{T}\cdot\UV{l}_\T{R}|\V{R}|^2-
\UV{l}_\T{T}\cdot\V{R}\,\,
\UV{l}_\T{R}\cdot\V{R}.
\label{eq:vector_identity}
\end{equation}
When the two lines are restricted to the same plane, the generalized 3D
expression \eqref{eq:3D_NDoF} reduces to the in-plane form
\eqref{eq:2D_mutual_shadow_length}.

The relation between this mutual-shadow (view) formulation and the optimal
parameterization approach developed in
\cite{bucci1998representation,pierri2020asymptotic,pierri2021ndf,leone2022dimension} is discussed in Sec.~\ref{subsec:formulation_3d_space}.
The generalized expression~\eqref{eq:3D_NDoF} for curves in \(\mathbb{R}^3\) is validated
numerically in Sec.~\ref{sec:DoF and sampling for curves in 3D}.

\subsection{One-DoF Sampling}

The proposed one-DoF sampling strategy is based on the interpretation that one spatial sample should represent one spatial DoF~\cite{li2026degreesfreedombeamforminglarge}. For a given transmitting and receiving geometry, the accumulated view length along the receiving line provides a natural coordinate for placing the samples. In this coordinate, adjacent sampling points are determined so that the cumulative view-length increment between them equals one wavelength, corresponding to one spatial DoF.

More specifically, let \(\xi\) denote the arc-length coordinate along the receiving line in Fig.~\ref{fig:research_problem}, the one-DoF sampling points \(\xi_n\) are selected so that
\begin{equation}
    L_\T{TR}(\xi_{n+1})-L_\T{TR}(\xi_n) = \lambda,
    \label{eq:lambda_domain}
\end{equation}
where \(L_\T{TR}(\xi)\) denotes the accumulated view-length~\eqref{eq:3D_NDoF} from the starting point of the receiving line to position \(\xi\). Equivalently, each interval between two adjacent samples contributes one unit to \(L_\T{TR}/\lambda\). Therefore, the physical spacing between samples is generally nonuniform, with more samples placed where the view length accumulates more rapidly along the receiving line.

To accurately represent the densely sampled channel operator, all curves are sampled on a sufficiently dense uniform grid. Specifically, a spacing of \(\lambda/50\), corresponding to fifty samples per wavelength, is adopted throughout and serves as the continuous-channel reference in all numerical examples. The singular values \(\sigma_n\) or \(\tilde{\sigma}_n\) are obtained by applying SVD to the channel matrix constructed from the scalar Green's function \(G(\V r_\T R,\V r_\T T)\) in~\eqref{eq:field_operator_general}. For a non-integer \(\mathcal N_{\T{DoF}}\), we use the ceiling
\(N=\lceil\mathcal N_{\T{DoF}}\rceil\) sampling points. The sampling intervals are adjusted over the whole receiving line, so that each cell represents approximately \(\mathcal N_{\T{DoF}}/N\) spatial DoF.

\begin{figure}[t!]
  \centering
  \includegraphics[width=1\linewidth]{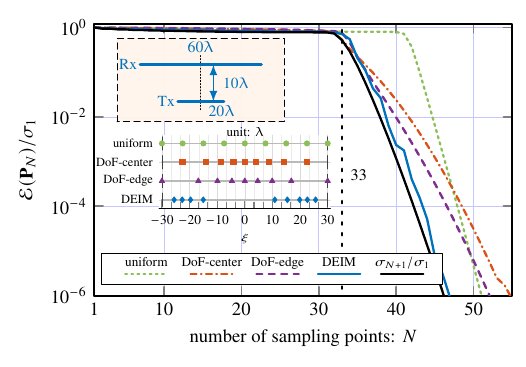}
  \caption{Sampling error~\eqref{eq:error_bound} using uniform sampling, DoF-center, DoF-edge and DEIM for the parallel-line configuration depicted in the inset. The inset below shows the sampling point distribution example on Rx when \(N=9\). The channel singular values \(\sigma_{N+1}/\sigma_1\) gives the dense-channel error bound~\eqref{eq:error_bound}.}
\label{fig:sampling_error}
\end{figure}

The total DoF estimate first provides a sampling criterion, namely that a one-DoF sampling set should contain at
least \(\lceil\mathcal N_{\T{DoF}}\rceil\) receiving points. This criterion, however, only specifies the minimum number of samples required and does not determine where the samples should be placed. Therefore, different sampling constructions may be used with the same DoF-level sampling budget. The question is then how to place the samples so that the sampled channel captures the dominant modal subspace and approaches the dense-channel residual bound.

A comparison of the sampling error for a parallel-line configuration is shown in Fig.~\ref{fig:sampling_error}. The vertical dotted line marks \(\lceil \mathcal N_{\T{DoF}}\rceil=33\), which separates the dominant propagating DoF region from the reactive post-knee modal tail. The NDoF is calculated using the closed-form expression for the parallel geometry in~\eqref{eq:parallel_shadow_f} given in App.~\ref{app:DoF_para_perp}. Uniform sampling places \(N\) receiving points at equal physical intervals over
\(\regR\), with the two endpoints included. DoF-center sampling divides the total DoF into \(N\) equal cells and places one point at the center of each cell. DoF-edge sampling places \(N\) points uniformly over the DoF interval, including its two endpoints. The sampling points are first placed at equal intervals over the DoF range, including both ends. These uniformly spaced DoF locations are then mapped back to the receiving region through the inverse of~\eqref{eq:lambda_domain}, yielding the corresponding physical sampling positions \(\xi_n\). For a symmetric receiving geometry with a symmetric DoF density, an odd \(N\) places one sample exactly at the center of the receiving line, whereas an even \(N\) places the two innermost samples symmetrically about the center. In contrast, DEIM selects \(N\) receiving points from the dense spatial grid based on the first \(N\) left singular vectors of the dense channel. The discrete empirical interpolation method (DEIM)~\cite{hochman2014reduced} curve serves as a matrix-based benchmark and also remains close to the DoF-based trends before the knee. 

For \(N<\lceil \mathcal N_{\T{DoF}}\rceil\), all sampling schemes exhibit relatively large errors because the number of samples is still insufficient to represent the dominant propagating modal subspace. After the NDoF knee, the difference between the sampling strategies becomes more pronounced.  Uniform sampling gives a relatively large error just after the knee, indicating that equal spacing in the physical coordinate does not capture the higher-order modal content efficiently in this transition region. However, its convergence improves for larger \(N\), since the increasing number of uniformly distributed samples gradually resolves the remaining spatial variations. The DoF-center sampling improves the convergence in the dominant-mode region, since each sample represents one DoF. However, it is less effective beyond the NDoF knee, where the remaining modes are more associated with boundary effects. In comparison, the DoF-edge rule gives a lower error after the transition region, because it includes the receiving-line boundaries from the beginning and therefore better captures the boundary-dominated modal components.

Among the considered sampling methods, DEIM gives the closest agreement with the dense-channel reference bound \(\sigma_{N+1}/\sigma_1\) in \eqref{eq:error_bound}. This is expected because DEIM selects sampling locations directly from the dominant singular-vector structure of the dense channel. However, this advantage comes at the cost of first constructing the dense channel matrix and evaluating its singular-value decomposition numerically.

The lower inset in Fig.~\ref{fig:sampling_error} illustrates the differences among the sampling-point distributions obtained by the considered methods for \(N=9\). Once the number of samples exceeds the estimated \(\mathcal N_{\T{DoF}}=33\), the DoF-edge and DEIM methods exhibit similar spatial distributions, with additional sampling points after the knee point preferentially allocated near the edges of the receiving region.

The DoF-edge method therefore provides a geometry-based alternative that approaches the DEIM trend more closely than DoF-center and uniform sampling in the reactive region, while avoiding a full matrix-based sample-selection procedure. This motivates a separation of the sampling rule into two parts. Before the DoF knee, the sample placement is mainly determined by the propagating DoF distribution, which describes the dominant modal content supported by the Tx--Rx geometry. After the knee, the remaining modes are more strongly affected by aperture truncation and boundary variations, and an additional remaining-mode density is needed to refine the sampling near the edges. The following subsections therefore introduce the propagating and reactive density components separately.

\subsection{Sampling Density for the Propagating Modes}
\label{subsec:formulation_3d_space}
The view-length~\eqref{eq:2D_NDoF} and~\eqref{eq:3D_NDoF} formulation describes the total number of spatial DoF. For a curved receiving line, it is also useful to know how these DoF are distributed along the Rx curve. We therefore introduce a local DoF density and first derive it for a line source and a receiving curve in \(\R^3\).

From \eqref{eq:3D_NDoF}, the view (mutual shadow) density can be obtained by collecting the contribution per unit length of the receiving line. It is written as
\begin{equation}
\ell_\T{TR}(\V r_\T R,\UV l_\T R)=
\int_{\regT}\frac{\big|(\UV l_\T T\times \V R)\cdot(\UV l_\T R\times \V R)
\big|}{|\V R|^3}\,\diff l_\T T .
\label{eq:mutual_shadow_density_3d}
\end{equation}
When the transmitting line is parameterized by its arc length \(\eta\), and
\(\V R(\eta)=\V r_\T T(\eta)-\V r_\T R,
\UV l_\T T(\eta)=\frac{\diff \V r_\T T(\eta)}{\diff \eta}\).
Then the view length density can be written as
\begin{equation}
\ell_\T{TR}(\V r_\T R,\UV l_\T R)
=\int_{\eta_-}^{\eta_+}
\frac{|\UV l_\T T(\eta)\times \V R(\eta)\cdot\UV l_\T R\times \V R(\eta)|}{|\V R(\eta)|^3}\,\diff \eta,
\label{eq:shadow_density_signed_kernel}
\end{equation}
where \(\eta_\pm\) denote the two endpoints
of the transmitting line. With projected direction function
\begin{equation}    
q(\eta;\V r_\T R,\UV l_\T R)={\UV l_\T R\cdot \V R(\eta)/
}{|\V R(\eta)|}
={\UV l_\T R\cdot \UV R(\eta)}
\end{equation}
and
\(\diff \V R(\eta)/\diff \eta=\UV l_\T T(\eta)\), direct
differentiation gives
\begin{equation}
\frac{\diff q}{\diff \eta}
=\frac{(\UV l_\T T\cdot\UV l_\T R)|\V R|^2-(\UV l_\T T\cdot\V R)
(\UV l_\T R\cdot\V R)}{
|\V R|^3},
\label{eq:q_derivative}
\end{equation}
where the last equality follows from the vector identity \eqref{eq:vector_identity}.

\begin{figure}[t!]
  \centering
  \includegraphics[width=1\linewidth]{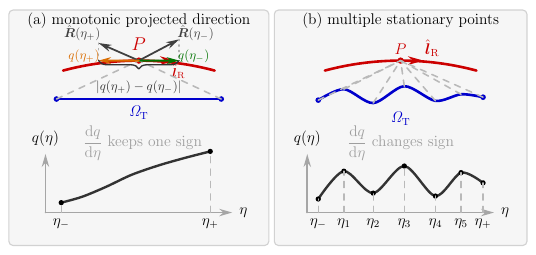}
\caption{Illustration of monotonic and sign-changing cases of the projected direction function \(q(\eta)\). Stationary points divide the transmitting region into sign-preserving intervals for evaluating the total variation.} \label{fig:sign_change_examples}
\end{figure}
If the derivative in \eqref{eq:q_derivative} does not change sign on
\([\eta_-,\eta_+]\), then \(q(\eta;\V r_\T R,\UV l_\T R)\) is monotonic over
the transmitting interval (see Fig.~\ref{fig:sign_change_examples}(a)). In this case, the absolute value can be taken outside the
integral, and
\begin{equation}
\ell_\T{TR}(\V r_\T R,\UV l_\T R)=\big|
q(\eta_+;\V r_\T R,\UV l_\T R)-q(\eta_-;\V r_\T R,\UV l_\T R)\big|.
\label{eq:conservative_shadow_density}
\end{equation}
Therefore, on a sign-preserving interval, the signed mutual shadow (view) density is conservative in the sense that its integral depends only on the endpoint values of \(q\).

If the derivative~\eqref{eq:q_derivative} changes sign inside the transmitting interval, the endpoint expression above is no longer sufficient because positive and negative contributions would otherwise cancel in the signed integral (see Fig.~\ref{fig:sign_change_examples}(b)). Let
\(\eta_1,\eta_2,\ldots,\eta_M\) denote the sign-changing points of \eqref{eq:q_derivative} inside \((\eta_-,\eta_+)\), and define
\(\eta_0=\eta_-, \eta_{M+1}=\eta_+\).
The mutual shadow (view) density should then be evaluated by summing the absolute contributions over the sign-preserving subintervals
\begin{equation}
\ell_\T{TR}(\V r_\T R,\UV l_\T R)
=\sum_{i=0}^{M}\left|q(\eta_{i+1};\V r_\T R,\UV l_\T R)-q(\eta_i;\V r_\T R,\UV l_\T R)\right|.
\label{eq:shadow_density_sign_change}
\end{equation}
Thus, even when \eqref{eq:q_derivative} changes sign, the density can still be computed from endpoint differences of \(q\), but the endpoints must include all turning points where the sign of \eqref{eq:q_derivative} changes.

The DoF density along the receiving line is\cite{gustafsson2025shadow}
\begin{equation}
\rho_\T p(\V r_\T R,\UV l_\T R)= \frac{1}{\lambda}
\ell_\T{TR}(\V r_\T R,\UV l_\T R),
\label{eq:dof_density_general_wire_3d}
\end{equation}
and the total DoF~\eqref{eq:2D_NDoF} is obtained by integrating this density~\eqref{eq:dof_density_general_wire_3d} over the receiving line
\begin{equation}
\mathcal N_\T{DoF}
= \int_{\regR}
\rho_\T p(\V r_\T R,\UV l_\T R)
\,\diff l_\T R =
\frac{L_\T{TR}}{\lambda}.
\label{eq:dof_from_density_general_wire_3d}
\end{equation}

The present approach is related to earlier nonredundant sampling and NDoF analyses~\cite{bucci1998representation,pierri2020asymptotic,pierri2021ndf,leone2022dimension}, in particular the reduced-field formulation of Bucci et al.~\cite{bucci1998representation}. In that formulation, the rapidly varying dominant propagation phase is first extracted from the electromagnetic field, and the remaining reduced field is treated as a spatially bandlimited function under a suitable optimal parameterization. Therefore, both approaches rely on the same physical principle that the relevant spatial variation is governed by the range of projected propagation directions seen from the source region. However, their objectives and resulting quantities are different. The reduced-field and related formulations mainly provide field-representation, interpolation, and spectral-dimension analyses. By contrast, the present formulation obtains the DoF density directly from the view length, or equivalently from the total variation of the projected direction function \(q(\eta)\). This gives an explicit geometric interpretation of local DoF accumulation, naturally accounts for sign changes through sign-preserving subintervals, and generates sampling locations from the cumulative DoF density. Moreover, the formulation extends directly to arbitrary receiving curves in \(3\T D\) without constructing the radiation operator~\cite{pierri2020asymptotic,leone2022dimension}, weighted adjoint~\cite{pierri2021ndf,leone2022dimension}, or lifting operator~\cite{leone2022dimension}.

\subsection{Sampling Density for the Reactive Modes}

Numerical sampling algorithms, such as DEIM used in Fig.~\ref{fig:sampling_error}, rely on the singular vectors of a densely sampled channel operator to identify sampling locations. These methods require the full channel matrix to be constructed in advance. In contrast, the goal of the present formulation is to obtain a geometry-based sampling density directly from the transmitting and receiving lines. Therefore, the singular-vector profiles are used here not as the sampling rule itself, but as a diagnostic tool for validating the proposed DoF-density interpretation. This comparison is also related to the edge effects studied in~\cite{brick2026interpreting}, where the modes beyond the propagating DoF region are shown to be more associated with boundary behavior.

\begin{figure}[t]
  \centering    
  \includegraphics[width=1\linewidth]{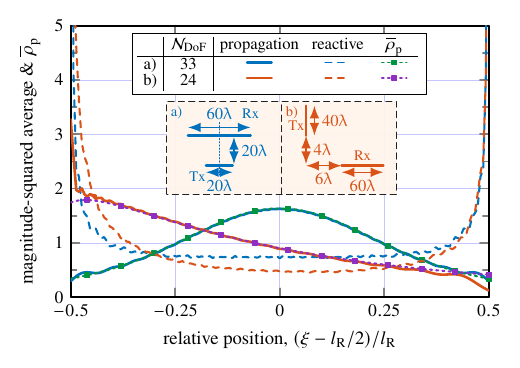}
\caption{Local singular-vector magnitude-squared averages and normalized DoF density for the parallel and perpendicular Tx--Rx geometries depicted in the inset. Solid curves denote the dominant propagating singular-vector modes, dashed curves denote the following 25 reactive modes, and dotted curves denote the normalized propagating DoF density \(\overline{\rho}_{\T p}\).}  \label{fig:density_singular_value_magnitude}
\end{figure}

Using the total-DoF relation in~\eqref{eq:dof_density_general_wire_3d}, the propagating DoF density is normalized by its spatial average as
\(\overline{\rho}_{\T p}(\xi)=\rho_{\T p}(\xi) l_{\T R}/\mathcal{N}_{\T{DoF}}\), where \(l_\T R\) is the length of the receiving line.
Fig.~\ref{fig:density_singular_value_magnitude} shows that \(\overline{\rho}_{\T p}(\xi)\) (see DoF density expressions~\eqref{eq:parallel_dL_final} and \eqref{eq:perp_dL_final} in App.~\ref{app:DoF_para_perp}) is closely aligned with the spatial distribution of the dominant propagating singular-vector modes. The corresponding NDoF values are obtained from the closed-form expressions in App.~\ref{app:DoF_para_perp}. In both the parallel and perpendicular geometries, the dotted density curves follow the solid curves, indicating that the view-length-based DoF density captures the spatial distribution of the dominant modal energy along the receiving line. This provides the basis for the propagating one-DoF sampling, where the samples are placed according to the cumulative integral of \(\overline{\rho}_{\T p}\).

The behavior changes beyond the propagating DoF region. Following the modal classification in~\cite{brick2026interpreting}, the singular-value spectrum can be divided into propagating, reactive, and noise-dominated regions. To obtain a representative reactive-mode profile while excluding the noise-dominated tail, the selected set is chosen to cover at least \(75\%\) of the identified reactive modes. Accordingly, the first \(25\) post-knee modes are considered here. These modes exhibit stronger concentration near the line boundaries, particularly around \(\xi/\ell_{\T R}=0\) and \(\xi/\ell_{\T R}=1\) as seen in Fig.~\ref{fig:density_singular_value_magnitude}. This stronger edge localization, together with possible rapid variations near stationary points of the projected-direction function, suggests that the post-knee modes are more associated with edge effects. Therefore, the edge variations of the dashed post-knee modes motivate an additional endpoint-enhanced reactive weighting, whose cumulative integral is used, in the same way as for the propagating density, to place the extra samples beyond the DoF limit.

The choice of an endpoint-enhanced weighting is motivated by the fact that boundary-dominated electromagnetic quantities often exhibit power-law behavior near edges and singular points. Classical edge-condition results show that surface quantities near conducting edges may exhibit integrable singularities
~\cite{meixner1972behavior,hurd1976edge}. In electrostatic conductor problems, this appears explicitly in the surface charge density near edges and corners, where charge-singularity analyses and conducting-disk solutions show power-law behavior with respect to the distance from the edge~\cite{morrison1976charge}. Similar power-law edge behavior is also commonly incorporated in singular basis functions in method of moments (MoM)~\cite{andersson1993moment, wilton2003incorporation}.
Although the post-knee singular modes considered here are not identical to static edge singularities, Fig.~\ref{fig:density_singular_value_magnitude}
shows that their spatial energy is enhanced near the line endpoints. 

To account for this behavior, we introduce an additional sampling density in a DoF-normalized coordinate. Let
\begin{equation}   
f(\xi)=\frac{1}{\mathcal N_\T{DoF}}
\int_{0}^{\xi}\rho_{\T p}(\xi)\,\diff \xi,
\label{eq:DoF_normalized_coordinate}
\end{equation}
where \(\rho_{\T p}(\xi)\) is the propagating DoF density in \eqref{eq:dof_density_general_wire_3d}. In this coordinate, the DoF sampling using~\eqref{eq:DoF_normalized_coordinate} becomes uniform. Motivated by the boundary-localized behavior of the reactive singular modes in Fig.~\ref{fig:density_singular_value_magnitude}, the reactive-mode density is chosen to enhance sampling near the line boundaries and is modeled by the following power-law form
\begin{equation} 
q_{\T r}(f)=f^{-p}+
(1-f)^{-p},
\end{equation}
where \(p\in(0,1)\). The purpose of applying this function is to improve the sampling density of the Rx edge. The corresponding reactive-mode density in the physical coordinate is defined as
\begin{equation}
\rho_{\T r}(\xi)=\frac{
q_{\T r}(f(\xi))\frac{\diff f}{\diff \xi}}{
\int_{0}^{1}q_{\T r}(f(\xi))\frac{\diff f}{\diff \xi}\diff \xi
}=\frac{1-p}{2\mathcal N_{\T{DoF}}}q_{\T r}(f(\xi))\rho_{\T p}(\xi).
\end{equation}
Therefore, after the propagating DoF knee \(N>\mathcal N_\T {DoF}\), the effective sampling density is written as
\begin{equation}
\rho_{\T {eff}}(\xi)=\rho_{\T p}(\xi)+(N-\mathcal N_\T{DoF})\rho_{\T r}(\xi),
\label{eq:reactive_density}
\end{equation} 
where \(\rho_{\T p}\) and \(\rho_{\T r}\) are normalized to integrate to \(\mathcal N_{\T{DoF}}\) and unity, respectively, ensuring that the resulting total density integrates to the prescribed number of sampling points \(N\). This construction preserves the geometry-aware propagating DoF distribution while adding extra samples in the regions where the reactive modes are expected to be most concentrated.

\begin{figure}[t]
  \centering
  \includegraphics[width=1\linewidth]{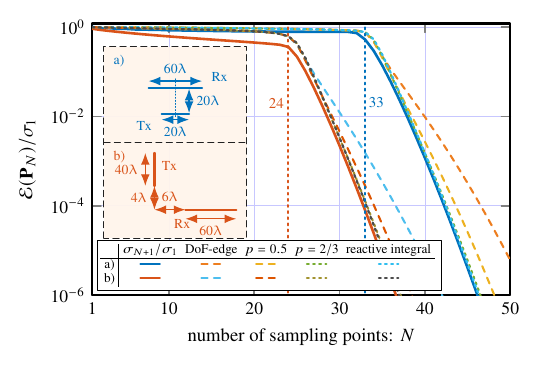}
\caption{Sampling-error comparison for the parallel- and perpendicular-line geometries depicted in the inset. The curves compare DoF-edge sampling, the proposed reactive-density
sampling in~\eqref{eq:reactive_density} with \(p=0.5\) and \(p=2/3\), and the reactive-integral benchmark derived from the averaged energy of
the 25 post-knee singular modes shown in
Fig.~\ref{fig:density_singular_value_magnitude}.}
\label{fig:sampling_error_reactive}
\end{figure}

The results in Fig.~\ref{fig:sampling_error_reactive} further confirm the need for the reactive-density correction beyond the propagating DoF limit. The vertical dotted lines mark the estimated DoF knees, with \(\lceil\mathcal N_{\T{DoF}}\rceil=33\) for case (a) and \(\lceil\mathcal N_{\T{DoF}}\rceil=24\) for case (b). Before these knees, the DoF-edge sampling follows the dense-channel residual trend reasonably well, indicating that the propagating DoF placement captures the dominant modal subspace. After the knees, however, the DoF-edge error decreases more slowly than \(\sigma_{N+1}/\sigma_1\), especially in the reactive tail. By adding the endpoint-enhanced reactive density~\eqref{eq:reactive_density}, the proposed sampling produces a faster error decay and follows the reference bound more closely. Among the tested choices, \(p=2/3\) provides the clearest improvement and is therefore used in all examples throughout this paper. This result indicates that the reactive-density term effectively compensates for the edge-localized modal behavior that is not fully captured by the propagating DoF placement alone. 

In addition, the numerical singular vectors themselves can indicate where the channel modes are spatially concentrated. When a dense reference channel is available, the local magnitude-squared averages of the corresponding singular vectors can be used as an density to assess where additional samples should be placed in the reactive region. This offers a useful reference for evaluating the proposed analytical correction and indicates that a more flexible reactive-density model may further improve the sampling performance beyond the simple power-law form~\eqref{eq:reactive_density}. The reactive-integral benchmark yields a further performance improvement because it derives the sampling density directly from the actual spatial energy distribution of the post-knee singular modes, rather than approximating their boundary-localized behavior using a prescribed power-law profile as seen in Fig.~\ref{fig:sampling_error_reactive}.

\section{Sampling for non-planar Tx--Rx geometries}
\label{sec:DoF and sampling for curves in 3D}
In this section, we provide representative numerical examples for different line configurations with Rx edges in 3D space to demonstrate the advantages of the proposed sampling method. For a general space curve embedded in \(3\T D\) space, the receiving region does not have a fixed global orientation. The DoF density is therefore determined by both the observation position and the local tangent direction. This geometry dependence affects not only the estimated NDoF, but also the sampling locations used to approximate the dominant channel subspace.

\begin{figure}[t!]
  \centering
  \includegraphics[width=1\linewidth]{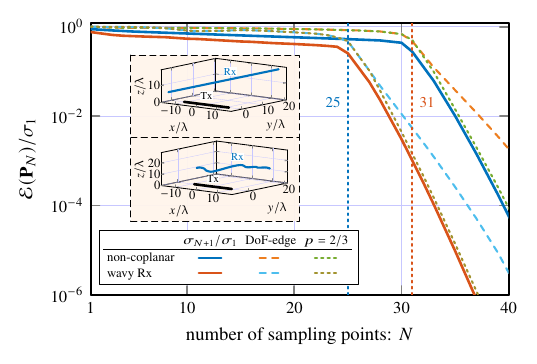}
\caption{Validation of the NDoF estimate~\eqref{eq:3D_NDoF} and the sampling error bound~\eqref{eq:error_bound} for two representative non-coplanar line configurations depicted in the inset. For the non-coplanar geometry, the Tx is a \(22\lambda\)-long straight line along the \(x\)-axis, while the Rx is parameterized as \(\V r_{\T R}(\xi)=\xi\UV u\), where \(\xi\in[-20\lambda,20\lambda]\) and \(\UV u\approx0.78\UV x+0.54\UV y+0.31\UV z\) is the unit direction vector of the Rx. For wavy Rx geometry, the Tx is a \(18\lambda\) straight line and the Rx is an open wavy curve embedded in 3D. The vertical dash lines indicate the estimated NDoF values~\eqref{eq:3D_NDoF}.}
\label{fig:error_dof}
\end{figure}

The first validation result is shown in Fig.~\ref{fig:error_dof}. The two examples represent non-coplanar receiving geometries with different levels of spatial variation. In both cases, the NDoF estimate~\eqref{eq:3D_NDoF} is close to the knee point of the singular-value spectrum \(\sigma_n/\sigma_1\), indicating that the propagation density gives a reasonable prediction of the number of dominant spatial modes. Compared with the DoF-edge sampling rule, the proposed sampling based on the effective density in~\eqref{eq:reactive_density} gives a faster error decay after the DoF transition and follows the lower bound \(\sigma_{N+1}/\sigma_1\) more closely.

The difference between the two sampling curves becomes most visible after the NDoF transition. The DoF-edge rule uses only the propagating DoF density, and therefore its error decreases more slowly once the post-knee modes start to contribute. By contrast, the effective-density sampling includes the reactive correction term and places additional samples in regions where the projected-direction function varies more rapidly. This improves the approximation of the post-knee modal subspace, especially for the wavy receiving curve where the local tangent direction changes along the line.

\begin{figure}[t!]
  \centering
  \includegraphics[width=1\linewidth]{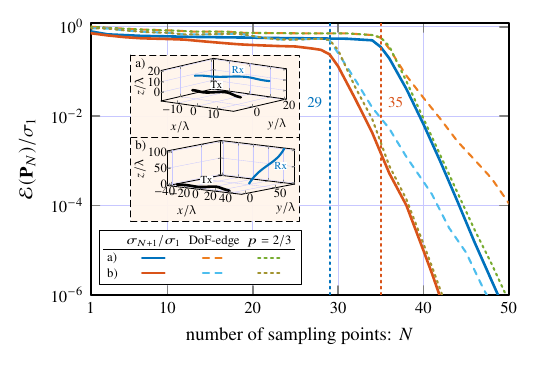}
\caption{Validation of the NDoF estimate~\eqref{eq:3D_NDoF} and the sampling-error bound~\eqref{eq:error_bound} for the two non-coplanar curved-line configurations shown in the insets. (a) consists of a gently curved Tx--Rx pair, while in (b) the longer Rx is located outside the positive-\(x\) edge of an arched Tx. The continuous NDoF estimate is rounded upward for DoF-edge and reactive-density sampling.}
\label{fig:error_dof_curved}
\end{figure}

Fig.~\ref{fig:error_dof_curved} further confirms that~\eqref{eq:3D_NDoF} remains applicable when both the transmitting and receiving regions are smoothly curved, provided that the curves do not exhibit excessive bending and their local tangent directions vary gradually. For both representative geometries, the predicted NDoF values are close to the transition regions of the corresponding sampling-error curves, indicating that the geometry-based expression captures the effective number of spatial modes supported by the curved Tx--Rx configuration. Beyond the estimated NDoF, the combined propagating and reactive-density sampling (\(p=2/3\)) remains closer to the singular-value lower bound than propagation-only DoF-edge sampling, demonstrating the accuracy of the NDoF estimate and the effectiveness of the proposed sampling strategy for moderately curved line regions.

\begin{figure}[t!]
  \centering
  \includegraphics[width=1\linewidth]{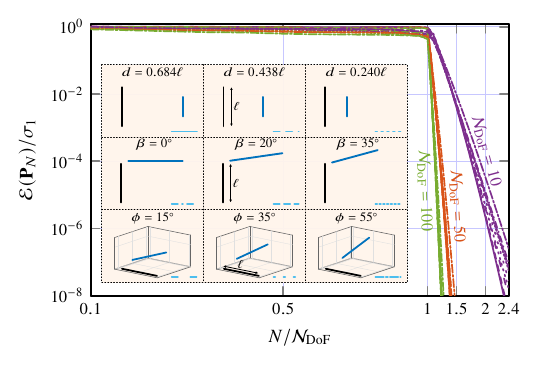}
\caption{Normalized sampling error of the proposed reactive-density method for the nine Tx--Rx line configurations shown in the inset. In the first row, a Tx of length \(\ell\) and an Rx of length \(0.5\ell\) are parallel, and \(d\) denotes their perpendicular separation. In the second row, the Tx is vertical and the Rx is placed completely outside its upper edge. The angle \(\beta\) denotes the in-plane rotation of the Rx relative to the horizontal direction shown in the inset, and the minimum horizontal and vertical clearances are both \(0.035\ell\). In the third row, the Tx has length \(\ell\), while the Rx has length \(20\ell/11\). The Rx center is displaced from the Tx center by \(4\ell/11\) and \(6\ell/11\) along the two transverse directions. The elevation angle is fixed at \(18.06^\circ\), whereas \(\phi\) denotes the azimuthal rotation of the Rx.}
\label{fig:normalized_error_dof}
\end{figure}

To compare sampling performance across geometries with different absolute numbers of spatial DoF, the sample count is normalized by the corresponding \(\mathcal N_{\T{DoF}}\). This normalization provides a common scale for assessing whether the error transition is governed primarily by the relative sampling level rather than by the specific Tx--Rx geometry. As shown in Fig.~\ref{fig:normalized_error_dof}, all nine configurations exhibit a clear error transition near \(N/\mathcal N_{\T{DoF}}=1\). 
For each prescribed NDoF, the curves corresponding to the parallel, rotated, and non-coplanar configurations remain relatively close after normalization. Although the detailed decay rates vary slightly with the Tx--Rx geometry, the transition consistently occurs around the normalized sampling ratio of one. This indicates that normalization by \(\mathcal N_{\T{DoF}}\) largely removes the geometry-dependent scaling and that the proposed reactive-density method automatically adapts the sampling locations to different relative positions and orientations.

The post-transition behavior also depends on the target NDoF. To facilitate a consistent comparison across different transmitting and receiving line configurations, the sampling errors are evaluated for a fixed \(\mathcal N_{\T {DoF}}\), with the wavelength determined from~\eqref{eq:2D_NDoF} as \(\lambda=L_{\T{TR}}/{\mathcal N_{\T {DoF}}}\). The curves for \(\mathcal N_{\T{DoF}}=100\) decrease more sharply and closer to \(N/\mathcal N_{\T{DoF}}=1\), whereas the cases with \(\mathcal N_{\T{DoF}}=10\) require a larger relative number of additional samples to achieve the same error level. This difference can be attributed to finite-sample and endpoint effects, which represent a larger fraction of the total sampling budget for systems with fewer supported modes. 

\section{Conclusions}
This paper presented a geometry-aware framework for near-field sampling based on the spatial DoF of electromagnetic propagation between transmitting and receiving line regions. Starting from a scalar line-source formulation and an inverse-source interpretation of the sampling problem, closed-form expressions were derived for the propagating-mode DoF density using the view-length relation. The cumulative DoF density was then used to construct one-DoF sampling locations, providing a direct geometric connection between the source--receiver configuration and the spatial distribution of sampling points. An operator-based sampling-error functional was also introduced and shown to be lower-bounded by the first neglected singular value of the continuous channel operator.

The theoretical development assumes scalar line sources, monotone smoothly varying transmitting and receiving curves with full mutual visibility, and propagation in homogeneous free space. Under these assumptions, the propagating DoF density accurately predicts both the total number of significant channel modes and their spatial distribution. Since the propagating density alone does not capture the localized behavior of higher-order modes beyond the singular-value-spectrum knee, an additional reactive-mode density was introduced to guide the placement of post-knee samples. Numerical examples for parallel, rotated, non-coplanar, and moderately curved line geometries demonstrated that the combined propagating--reactive sampling strategy closely reproduces the singular-value spectrum and consistently outperforms sampling based solely on the propagating-mode density.

Although the present work focuses on scalar line-source systems, the underlying framework naturally extends to vector electromagnetic fields by incorporating multiple current and field components. More generally, the geometric interpretation provided by the view-length formulation suggests that the proposed DoF-density concept is not limited to line geometries. Future work will investigate extensions to closed loops, arbitrary space curves, and two-dimensional surfaces, where the DoF density is expected to become a surface quantity governing both the number and distribution of optimal sampling points. Such developments could establish a unified geometric framework linking electromagnetic degrees of freedom, sampling, resolution, beamforming, and inverse-source reconstruction for general radiating structures.

\appendices

\section{NDoF and DoF Density}\label{app:DoF_para_perp}
\subsubsection{Parallel Geometry}
Consider a transmitting line source of length $\ell_1$ and a receiving line segment of length $\ell_2$. The two line segments are parallel, separated by a vertical distance $d_2$, and the center of the receiving segment is shifted by $d_1$ along the $x$-direction (see inset in Fig.~\ref{fig:density_singular_value_magnitude}). 

For this parallel configuration, the NDoF can be written as~\cite{li2026degreesfreedombeamforminglarge}
\begin{equation}
\mathcal N_{\T{DoF}}^{\parallel}
=[f(\delta_+)-f(\delta_-)]/\lambda,
\label{eq:parallel_shadow_f}
\end{equation}
where $\delta_\pm = |\ell_1 \pm \ell_2|,$ and
\begin{equation}
f(\delta)=\sqrt{d_2^2+\left({\delta}/{2}-d_1\right)^2
}+\sqrt{d_2^2+\left({\delta}/{2}+d_1\right)^2}.
\label{eq:f_delta_parallel}
\end{equation}
When the receiving line becomes differential, i.e., \(\ell_2=\diff\ell\), the mutual shadow length $\diff L_{\T{TR}}^{\parallel}$ becomes
\begin{equation}
\diff L_{\T{TR}}^{\parallel}=f(\ell_1+\diff\ell)
-f(\ell_1-\diff\ell)\approx2f'(\ell_1)\diff\ell,
\label{eq:parallel_differential_start}
\end{equation}
which using the first-order Taylor expansion around $\ell_1$, $f(\ell_1\pm\diff\ell)\approx f(\ell_1)\pm f'(\ell_1)\diff\ell.$
Thus, the DoF density is
\begin{equation}
\rho_{\T{p}}^{\parallel}=\frac{\frac{\ell_1}{2}+d_1}{\sqrt{d_2^2+\left(\frac{\ell_1}{2}+d_1\right)^2}}-
\frac{d_1-\frac{\ell_1}{2}}
{\sqrt{d_2^2+\left(\frac{\ell_1}{2}-d_1\right)^2}}.
\label{eq:parallel_dL_final}
\end{equation}

\subsubsection{Perpendicular Geometry}

For the perpendicular configuration, let the transmitting line occupy the interval from $x=d_1$ to $x=d_1+\ell_1$, while the receiving line is vertical. The lower and upper endpoints of the receiving line are
$z_1=d_2, z_2=d_2+\ell_2$. The NDoF is written as ~\cite{li2026degreesfreedombeamforminglarge}
\begin{equation}
\mathcal N_{\T{DoF}}^{\perp}=[F(z_1)-F(z_2)]/\lambda,
\label{eq:perp_shadow_F}
\end{equation}
where $F(z)=\sqrt{z^2+(\ell_1+d_1)^2}-\sqrt{z^2+d_1^2}$.

If the receiving line becomes differential, $\ell_2=\diff\ell,$ then $z_2=z_1+\diff\ell$.
Therefore, mutual shadow length $ L_{\T{TR}}^{\perp}=F(z_1)-F(z_1+\diff\ell)$. Using the Taylor expansion
$F(z_1+\diff\ell)\approx F(z_1)+F'(z_1)\diff\ell,$
we obtain $\diff L_{\T{TR}}^{\perp}\approx-F'(z_1)\diff\ell$. Thus, the DoF density $\rho_{\T{p}}^{\perp}$ is
\begin{equation}
\rho_{\T{p}}^{\perp}={d_2}/{\sqrt{d_2^2+d_1^2}}-
{d_2}/{\sqrt{d_2^2+(\ell_1+d_1)^2}}.
\label{eq:perp_dL_final}
\end{equation}

\bibliographystyle{IEEEtran}
\bibliography{citations}

@article{pierri2020asymptotic,
  title={Asymptotic study of the radiation operator for the strip current in near zone},
  author={Pierri, Rocco and Moretta, Raffaele},
  journal={Electronics},
  volume={9},
  number={6},
  pages={911},
  year={May, 2020},
  publisher={MDPI}
}

@article{pierri2021ndf,
  title={{NDF} of the near-zone field on a line perpendicular to the source},
  author={Pierri, Rocco and Moretta, Raffaele},
  journal={IEEE Access},
  volume={9},
  pages={91649--91660},
  year={Jun. 2021},
  publisher={IEEE}
}

@article{leone2022dimension,
  title={Dimension and sampling of the near-field and its intensity over curves},
  author={Leone, Giovanni and Moretta, Raffaele and Pierri, Rocco},
  journal={IEEE Open J. Antennas Propag.},
  volume={3},
  pages={412--424},
  year={Feb. 2022},
  publisher={IEEE}
}

@article{gustafsson2025shadow,
  title={Shadow area and degrees of freedom for free-space communication},
  author={Gustafsson, Mats},
  journal={IEEE J. Sel. Areas Inf. Theory},
  volume  = {6},
  pages   = {325-337},
  year={2025},
  publisher={IEEE}
}

@article{gustafsson2025degrees,
  title={Degrees of freedom for radiating systems},
  author={Gustafsson, Mats},
  journal={IEEE Trans. Antennas Propag.},
  volume={73},
  number={2},
  pages={1028--1038},
  year={Feb. 2025},
  publisher={IEEE}
}

@book{weyl1911asymptotische,
  author    = {Arendt, W. and Nittka, R. and Peter, W. and Steiner, F.},
  title     = {Weyl's Law: Spectral Properties of the Laplacian in Mathematics and Physics},
  publisher = {John Wiley \& Sons, Ltd},
  year      = {2009},
  chapter   = {1},
  pages     = {1--71}
}

@article{miller2000communicating,
  author  = {Miller, David AB},
  title   = {Communicating with Waves Between Volumes: Evaluating Orthogonal Spatial Channels and Limits on Coupling Strengths},
  journal = {Appl. Opt.},
  volume  = {39},
  number  = {11},
  pages   = {1681--1699},
  year    = {Apr. 2000},
  doi     = {10.1364/AO.39.001681}
}

@article{jensen2008capacity,
  author  = {Jensen, M. A. and Wallace, J. W.},
  title   = {Capacity of the Continuous-Space Electromagnetic Channel},
  journal = {IEEE Trans. Antennas Propag.},
  volume  = {56},
  number  = {2},
  pages   = {524--531},
  year    = {Feb. 2008},
  doi     = {10.1109/TAP.2007.915421}
}

@article{bucci1989degrees,
  author  = {Bucci, O. M. and Franceschetti, G.},
  title   = {On the Degrees of Freedom of Scattered Fields},
  journal = {IEEE Trans. Antennas Propag.},
  volume  = {37},
  number  = {7},
  pages   = {918--926},
  year    = {Jul. 1989},
  doi     = {10.1109/8.29386}
}

@article{maisto2021near,
  title={Near-field transverse resolution in planar source reconstructions},
  author={Maisto, Maria Antonia and Pierri, Rocco and Solimene, Raffaele},
  journal={IEEE Trans. Antennas Propag.},
  volume={69},
  number={8},
  pages={4836--4845},
  year={Aug. 2021},
  publisher={IEEE}
}

@article{puggelli2025maximizing,
  title={Maximizing independent channels and efficiency in {BTS} array antennas via {EM} degrees of freedom},
  author={Puggelli, Federico and Biscontini, Bruno and Martini, Enrica and Maci, Stefano},
  journal={IEEE Trans. Antennas Propag.},
  year={Jun. 2025},
  volume={73},
  number={6},
  pages={3444-3458},
  publisher={IEEE}
}

@ARTICLE{Bucci2025,
  author={Bucci, Ovidio Mario and Migliore, Marco Donald},
  journal={IEEE Antennas Propag. Mag.}, 
  title={Degrees of Freedom and Sampling Representation of Electromagnetic Fields: Concepts and applications.}, 
  year={Jun. 2025},
  volume={67},
  number={3},
  pages={10-22},
  doi={10.1109/MAP.2024.3513216}}

@article{bucci1998representation,
  author  = {Bucci, O. M. and Gennarelli, C. and Savarese, C.},
  title   = {Representation of Electromagnetic Fields over Arbitrary Surfaces by a Finite and Nonredundant Number of Samples},
  journal = {IEEE Trans. Antennas Propag.},
  volume  = {46},
  number  = {3},
  pages   = {351--359},
  year    = {Mar. 1998},
  doi     = {10.1109/8.662654}
}

@book{franceschetti2017wave,
  author    = {Franceschetti, M.},
  title     = {Wave Theory of Information},
  publisher = {Cambridge Univ. Press},
  address   = {Cambridge, U.K.},
  year      = {2017}
}

@article{dardari2020communicating,
  title={Communicating with large intelligent surfaces: Fundamental limits and models},
  author={Dardari, Davide},
  journal={IEEE J. Sel. Areas Commun.},
  volume={38},
  number={11},
  pages={2526--2537},
  year={Nov. 2020},
  publisher={IEEE}
}

@article{yuan2024breaking,
  author={Yuan et al., Shuai S. A.},
  title   = {Breaking the Degrees-of-Freedom Limit of Holographic {MIMO} Communications: A {3-D} Antenna Array Topology},
  journal = {IEEE Trans.
Veh. Technol.},
  volume  = {73},
  number  = {8},
  pages   = {11276--11288},
  year    = {Aug. 2024},
}

@article{maisto2021efficient,
  author  = {Maisto, M. A. and Leone, G. and Brancaccio, A. and Solimene, R.},
  title   = {Efficient Planar Near-Field Measurements for Radiation Pattern Evaluation by a Warping Strategy},
  journal = {IEEE Access},
  volume  = {9},
  pages   = {62255--62265},
  year    = {Apr. 2021},
  doi     = {10.1109/ACCESS.2021.3074786}
}

@article{hochman2014reduced,
  title={Reduced-order models for electromagnetic scattering problems},
  author={Hochman, Amit and Villena, Jorge Fernández and Polimeridis, Athanasios G. and Silveira, Luís Miguel and White, Jacob K. and Daniel, Luca},
  journal={IEEE Trans. Antennas Propag.},
  volume={62},
  number={6},
  pages={3150--3162},
  year={Jun. 2014},
  publisher={IEEE}
}

@article{poon2005degrees,
  title={Degrees of freedom in multiple-antenna channels: A signal space approach},
  author={Poon, Ada SY and Brodersen, Robert W and Tse, David NC},
  journal={IEEE Trans. Inf.
Theory},
  volume={51},
  number={2},
  pages={523--536},
  year={Feb. 2005},
  publisher={IEEE}
}

@article{li2026degreesfreedombeamforminglarge,
  title={Degrees of Freedom and Beamforming for Large Intelligent Surfaces},
  author={Li, Jiawang and Saberkari, Alireza and Lau, Buon Kiong and Gustafsson, Mats},
  journal={arXiv preprint arXiv:2606.19666},
  year={2026}
}

@book{harrington1993field,
  title={Field Computation by Moment Methods},
  author={Harrington, Roger F},
  year={1968},
  publisher={New York, NY: Macmillan}
}

@article{eckart1936approximation,
  title={The approximation of one matrix by another of lower rank},
  author={Eckart, Carl and Young, Gale},
  journal={Psychometrika},
  volume={1},
  number={3},
  pages={211--218},
  year={1936},
  publisher={Springer-Verlag}
}

@book{ben2003generalized,
  title={Generalized Inverses: Theory and Applications},
  author={Ben-Israel, Adi and Greville, Thomas},
  year={1974},
  publisher={New York, NY, USA: Wiley}
}

@article{brick2026interpreting,
  title={Interpreting Moment Matrix Blocks Spectra using Mutual Shadow Area},
  author={Brick, Yaniv and Andriulli, Francesco P and Gustafsson, Mats},
  journal={IEEE Trans. Antennas Propag.},
  year={2026},
  volume={},
  number={},
  pages={1-1},
}

@article{andersson1993moment,
  title={Moment-method calculations on apertures using basis singular functions},
  author={Andersson, Tommy},
  journal={IEEE Trans. Antennas Propag.},
  volume={41},
  number={12},
  pages={1709--1716},
  year={Dec. 1993},
  publisher={IEEE}
}

@article{wilton2003incorporation,
  title={Incorporation of edge conditions in moment method solutions},
  author={Wilton, D and Govind, S},
  journal={IEEE Trans. Antennas Propag.},
  volume={25},
  number={6},
  pages={845--850},
  year={Nov. 2003},
  publisher={IEEE}
}

@article{morrison1976charge,
  title={Charge singularity at the corner of a flat plate},
  author={Morrison, Jo A and Lewis, JA},
  journal={SIAM J. Appl. Math.},
  volume={31},
  number={2},
  pages={233--250},
  year={1976},
  publisher={SIAM}
}

@article{hurd1976edge,
  title={The edge condition in electromagnetics},
  author={Hurd, R},
  journal={IEEE Trans. Antennas Propag.},
  volume={24},
  number={1},
  pages={70--73},
  year={Jan. 1976},
  publisher={IEEE}
}

@article{meixner1972behavior,
  title={The behavior of electromagnetic fields at edges},
  author={Meixner, Josef},
  journal={IEEE Trans. Antennas Propag.},
  volume={20},
  number={4},
  pages={442--446},
  year={Jul. 1972},
  publisher={IEEE}
}

@article{qureshi2012efficient,
  title={Efficient near-field far-field transformation for nonredundant sampling representation on arbitrary surfaces in near-field antenna measurements},
  author={Qureshi, M Ayyaz and Schmidt, Carsten H and Eibert, Thomas F},
  journal={IEEE Trans. Antennas Propag.},
  volume={61},
  number={4},
  pages={2025--2033},
  year={Dec. 2012},
  publisher={IEEE}
}

@article{mezieres2021antenna,
  title={Antenna characterization from a small number of far-field measurements via reduced-order models},
  author={M{\'e}zi{\`e}res, Nicolas and Mattes, Michael and Fuchs, Benjamin},
  journal={IEEE Trans. Antennas Propag.},
  volume={70},
  number={4},
  pages={2422--2430},
  year={Oct. 2021},
  publisher={IEEE}
}

@article{fuchs2017fast,
  title={Fast antenna far-field characterization via sparse spherical harmonic expansion},
  author={Fuchs, Benjamin and Le Coq, Laurent and Rondineau, S{\'e}bastien and Migliore, Marco Donald},
  journal={IEEE Trans. Antennas Propag.},
  volume={65},
  number={10},
  pages={5503--5510},
  year={Aug. 2017},
  publisher={IEEE}
}

@article{capozzoli2016singular,
  title={Singular value optimization in inverse electromagnetic scattering},
  author={Capozzoli, Amedeo and Curcio, Claudio and Liseno, Angelo},
  journal={IEEE Antennas Wireless Propag. Lett.},
  volume={16},
  pages={1094--1097},
  year={Dec. 2016},
  publisher={IEEE}
}

@article{randazzo2021two,
  author    = {Randazzo, Andrea and others},
  title     = {A Two-Step Inverse-Scattering Technique in Variable-Exponent Lebesgue Spaces for Through-the-Wall Microwave Imaging: Experimental Results},
  journal   = {IEEE Trans. Geosci. Remote Sens.},
  volume    = {59},
  number    = {9},
  pages     = {7189--7200},
  year      = {Sep. 2021},
  publisher = {IEEE}
}

@article{marengo2000inverse,
  title={Inverse source problem and minimum-energy sources},
  author={Marengo, Edwin A and Devaney, Anthony J and Ziolkowski, Richard W},
  journal={J. Opt. Soc. Amer. A},
  volume={17},
  number={1},
  pages={34--45},
  year={2000},
  publisher={Optical Society of America}
}

@article{yaghjian1986overview,
  title={An overview of near-field antenna measurements},
  author={Yaghjian, Arthur},
  journal={IEEE Trans. Antennas Propag.},
  volume={34},
  number={1},
  pages={30--45},
  year={Jan. 1986},
  publisher={IEEE}
}

@article{solimene2019sampling,
  title={Sampling approach for singular system computation of a radiation operator},
  author={Solimene, Raffaele and Maisto, Maria Antonia and Pierri, Rocco},
  journal={J. Opt. Soc. Amer. A.},
  volume={36},
  number={3},
  pages={353--361},
  year={2019},
  publisher={Optical Society of America}
}

@article{han2026reactive,
  title={Reactive Near-Field Reconstruction via {R}ayleigh--{S}ommerfeld Integral: Ill-Conditioning Analysis and Regularization},
  author={Han, Dong-Hao and Wei, Xing-Chang},
  journal={IEEE Trans. Microw. Theory Techn.},
  year={2026},
  pages={1-10},
  publisher={IEEE}
}

@article{solimene2018inverse,
  title={Inverse source in the near field: {T}he case of a strip current},
  author={Solimene, Raffaele and Maisto, Maria Antonia and Pierri, Rocco},
  journal={J. Opt. Soc. Amer. A.},
  volume={35},
  number={5},
  pages={755--763},
  year={2018},
  publisher={Optical Society of America}
}

@article{candes2008introduction,
  title={An introduction to compressive sampling},
  author={Cand{\`e}s, Emmanuel J and Wakin, Michael B},
  journal={IEEE Signal Process. Mag.},
  volume={25},
  number={2},
  pages={21--30},
  year={Mar. 2008},
  publisher={IEEE}
}

@article{bangun2022optimizing,
  title={Optimizing sensing matrices for spherical near-field antenna measurements},
  author={Bangun, Arya and Culotta-L{\'o}pez, Cosme},
  journal={IEEE Trans. Antennas Propag.},
  volume={71},
  number={2},
  pages={1716--1724},
  year={Dec, 2022},
  publisher={IEEE}
}

@book{marks1993advanced,
  author    = {Marks II, Robert J.},
  title     = {Advanced Topics in Shannon Sampling and Interpolation Theory},
  address   = {New York, NY, USA},
  publisher = {Springer-Verlag},
  year      = {1993}
}

@incollection{gabor1961light,
  author    = {Gabor, Dennis},
  title     = {{IV} Light and Information},
  booktitle = {Progress in Optics{\normalfont, vol.~1}},
  address   = {Amsterdam, The Netherlands},
  publisher = {Elsevier},
  year      = {1961},
  pages     = {109--153}
}

@article{miller2019waves,
  title={Waves, modes, communications, and optics: A tutorial},
  author={Miller, David AB},
  journal={Adv. Opt. Photon.},
  volume={11},
  number={3},
  pages={679--825},
  year={2019},
  publisher={Optical Society of America}
}

@article{piestun2000electromagnetic,
  title={Electromagnetic degrees of freedom of an optical system},
  author={Piestun, Rafael and Miller, David AB},
  journal={JOSA A},
  volume={17},
  number={5},
  pages={892--902},
  year={2000},
  publisher={OSA}
}

@article{bucci1997electromagnetic,
  title={Electromagnetic inverse scattering: Retrievable information and measurement strategies},
  author={Bucci, O. M. and Isernia, T},
  journal={Radio Sci.},
  volume={32},
  number={6},
  pages={2123--2137},
  year={1997},
  publisher={AGU}
}

@article{migliore2025intuitive,
  title={An Intuitive Approach to the Optimal Sampling of an Electromagnetic Field},
  author={Migliore, Marco Donald},
  journal={Sensors},
  volume={25},
  number={24},
  pages={7591},
  year={2025},
  publisher={MDPI}
}

\vfill

\end{document}